\documentclass[namedreferences]{solarphysics}

\usepackage[hyperref,optionalrh,showbiblabels]{spr-sola-addons} 
\usepackage[optionalrh]{spr-sola-addons} 
\usepackage{epsfig}          
\usepackage{graphicx}        
\usepackage{courier}         
\usepackage{amssymb}        
\usepackage{color}           
\usepackage{breakurl}        

\newcommand{\etal}{{\it et al.}}

\newcommand{\aap}{    {\it Astron. Astrophys.}}
\newcommand{\aaps}{   {\it Astron. Astrophys. Suppl.}}

\newcommand{\apj}{    {\it Astrophys. J.}}

\newcommand{\solphys}{{\it Solar Phys.}}

\newcommand{\na}{ {\it New Astronomy.}}

\chardef\us=`\_

\begin{document}

\begin{article}
\begin{opening}

\title{Precise Reduction of Solar Spectra Observed by the 1-meter \emph{New Vacuum Solar Telescope}}

\author[addressref={aff1,aff2},email={cyf2012@ynao.ac.cn}]{\inits{Y.F.}\fnm{Yunfang}~\lnm{Cai}~\orcid{0000-0002-4956-4320}}

\author[addressref=aff1,email={xuzhi@ynao.ac.cn}]{\inits{Z.}\fnm{Zhi}~\lnm{Xu}~\orcid{0000-0001-7084-3511}}

\author[addressref=aff1,email={lizg@ynao.ac.cn}]{\inits{Z.G.}\fnm{Zhenggang}~\lnm{Li}~\orcid{0000-0002-4635-984X}}

\author[addressref=aff1,email={xiangyy@ynao.ac.cn}]{\inits{Y.Y.}\fnm{Yongyuan}~\lnm{Xiang}~\orcid{0000-0002-5261-6523}}

\author[addressref={aff1,aff2},email={cyc@ynao.ac.cn}]{\inits{Y.C.}\fnm{Yuchao}~\lnm{Chen}~\orcid{0000-0001-8279-7014}}

\author[addressref=aff1,email={fiona@ynao.ac.cn}]{\inits{Y.}\fnm{Yu}~\lnm{Fu}~\orcid{0000-0002-1570-1198}}

\author[addressref=aff1,corref,email={jkf@ynao.ac.cn}]{\inits{K.F.}\fnm{Kaifan}~\lnm{Ji}~\orcid{0000-0001-8950-3875}}

\address[id=aff1]{Yunnan Observatories, Chinese Academy of Sciences, Kunming 650011, China}
\address[id=aff2]{University of Chinese Academy of Sciences, Beijing 100049, China}

\runningauthor{Y.F. Cai \etal}
\runningtitle{Precise Reduction of Solar Spectra}

\begin{abstract}

We present a precise and complete procedure for processing spectral data observed by the 1-meter \emph{New Vacuum Solar Telescope} (NVST). The procedure is suitable for both the sit-and-stare and raster-scan spectra. In this work, the geometric distortions of the spectra are firstly corrected for subsequent processes. Then, considering the temporal changes and the remnants of spectral lines in the flat-field, the original flat-field matrix is split into four independent components for ensuring a high precision flat-fielding correction, consisting of the continuum gradient matrix, slit non-uniform matrix, CCD dust matrix, and interference fringe matrix. Subsequently, the spectral line drifts and intensity fluctuations of the science data are further corrected. After precise reduction with this procedure, the measuring accuracies of the Doppler velocities for different spectral lines and of the oscillation curves of the chromosphere and photosphere are measured. The results show that the highest measuring accuracy of the Doppler velocity is within 100 m\,s\textsuperscript{-1}, which indicates that the characteristics of the photosphere and chromosphere can be studied co-spatially and co-temporally with the reduced spectra of NVST.
\end{abstract}


\end{opening}

\section{Introduction}
     \label{S-1}

The solar grating spectrometer is one of the most classical equipments used to obtain solar spectral information. At present, many famous solar telescopes are equipped with grating spectrometers, such as the \emph{GREGOR Infrared Spectrograph} (GRIS) \citep{Collados2012}, and the \emph{Fast Imaging Solar Spectrograph} (FISS) of the \emph{New Solar Telescope} (NST) (\citealp{Chae2013}).
The \emph{New Vacuum Solar Telescope} (NVST) of China (\citealp{Liu2014}) was built and put into observation in 2010, with two spectrometers which are also grating spectrometers.

As is well known, the spectral observations taken by the grating spectrometer are always subject to the optical system of the telescope and the cameras used. To acquire physical parameters of the solar atmosphere accurately, the raw spectrum data requires post-processing.
In the earlier years, the \emph{Swedish Vacuum Solar Telescope} (SVST) team proposed some approaches to solve the basic problems encountered by the grating spectrograms \citep{Johannesson1992,Kiselman1994}. After that, based on these approaches, different improvements were introduced according to the different problems existed in different spectrometers.
For example, the \emph{Vacuum Tower Telescope} (VTT) team separates the slit non-uniformity from the flat-field pattern (\citealp{VTT2002}) to eliminate the influence of the spectrograph drift and temporal changing of the flat-field. The NST team acquires several groups of flat data (\citealp{Chae2013}) in which the position of the spectral line on the detector is changed by setting the different grating angles. They use those groups of flat data to demodulate the final flat-field matrix (\citealp{Chae2004}). The NVST team proposed a set of reduction procedure (\citealp{wang2013}), which is suitable for processing the science data whose acquisition time is very close to the flat data. For the spectrums acquired in long time series of several hours, the time-varying factors should be considered. In addition, the remnants of spectral lines always exist in the obtained flat-field and bring new errors for processing science data. In this paper, we propose a new procedure to solve these problems and to achieve higher processing accuracy.

The basic purpose of spectrum reduction is to achieve high accuracy measurements of physical parameters of the solar atmosphere. The Doppler velocity is one of the most important parameters, which is greatly different at different solar atmosphere heights. For example, the average Doppler velocity in the chromosphere is above 1 km\,s\textsuperscript{-1} \citep{Yang2013,Hong2014}, while in the photosphere it is about 0.1\,--\,0.4 km\,s\textsuperscript{-1}. The NVST multi-band spectrometer has a dispersion power of about 130,000 in the H$\alpha$ band, and the spectral sampling rate is about 0.024\,{\AA} pixel\textsuperscript{-1}, so each pixel corresponds to 1.1 km\,s\textsuperscript{-1} Doppler velocity. To study the characteristics both of the chromosphere and photosphere simultaneously, the measuring accuracy of the Doppler velocity is required to reach the level same as that of photosphere, namely up to the one-tenth pixel level.

The NVST and its spectrometers are introduced in Section~\ref{S-2}, and the observations are described in Section~\ref{S-3}. The detailed processing methods are exhibited in Section~\ref{S-4}. The reduced results and the measuring accuracy of the Doppler velocity are given in Section~\ref{S-5}, and the conclusion is presented in Section~\ref{S-6}.

\section{Instruments}
\label{S-2}

The NVST (\citealp{Liu2014}) is a vacuum solar telescope with a 985 mm clear aperture. Its scientific goal is to obtain solar information by high resolution imaging (\citealp{Xiang2016}) and spectral observations in the wavelength ranges from 0.3 to 2.5\,$\mu$m. There are three main groups of instruments, the 151-component adaptive optics (AO) system \citep{1Rao2016,2Rao2016} and polarization analyzer (PA), the high resolution imaging system and two vertical grating spectrometers: the \emph{Multi-Band Spectrometer} (MBS) and the \emph{High Dispersion Spectrometer} (HDS).

The MBS uses a blazed grating working in the visible bands with 6\,m focal length, and the HDS uses an echelle grating working in the near-infrared bands with 9\,m focal length. They share the same slit but cannot work at same time, because their structures are perpendicular to each other. To switch between the two spectrometers, the collimators should be changed, and the slit direction should be rotated. In addition, the spectrometers are equipped with a slit-jaw imaging system and a scanning mechanism to do raster-scan observations.
The MBS can cover any part of the visible wavebands simultaneously, but only three bands, the H$\alpha$ 6562.8\,{\AA} band, Ca {\sc II} 8542\,{\AA} band and Fe {\sc I} 5324\,{\AA} band, are in daily observation. The basic parameters of MBS are listed in Table~\ref{T-1}.

\begin{table}
\caption{The main performance parameters of MBS: incident angle=$33.5^{\circ}$, blaze angle=$36.8^{\circ}$, grating constant: 1200 grooves\,mm\textsuperscript{-1}, slit width=60\,$\mu$m (0.3$''$).}
\label{T-1}
\begin{tabular}{lccccc}     
  \hline                   
     & grating order & diffraction angle~  & linear dispersion &  sampling rate \\
     &               & ($^{\circ}$)        & (mm/{\AA})        & (m{\AA}/pixel) \\
  \hline
H$\alpha$ (6562.8\,{\AA})  & 1 & 13.6  & 0.74 & 12 \\
Ca {\sc II} (8542\,{\AA}) & 1 & 28.2  & 0.82 & 11 \\
Fe {\sc I} (5324\,{\AA})  & 2 & 46.5  & 2.0  & 4.3 \\
  \hline
\end{tabular}
\end{table}

\section{Observations} 
      \label{S-3}

The data used in this paper were observed by the MBS in the H$\alpha$ band with the aid of the AO system on 2 March 2016. The width of incident slit is 60\,$\mu$m, corresponding to 0.3$''$. The camera used is the PCO4000. Its original sensor size is 4008 $\times$ 2672 pixels$^{2}$, and the pixel size is 9\,$\mu$m. In the observation, a binning factor of 2 $\times$ 4 is used. Thus the size of the observation data is 2004 $\times$ 668 pixels$^{2}$.
The H$\alpha$ 6562.8\,{\AA} line is observed in the 1st order. Therefore, the spectral sampling rate is about 0.024\,{\AA} pixel\textsuperscript{-1} and the spatial sampling rate is 0.164$''$ pixel\textsuperscript{-1}.

Three kinds of observation data are acquired: dark data, flat data, and science data. The \emph{X} and \emph{Y} axes of the data denote the dispersion and space directions, respectively. The dark data are acquired when the dome is closed. The integration time was 60\,ms, the same as that of the flat data and science data. One hundred frames of dark data are acquired continuously and only the average frame $\emph{D}_{\rm mean}$ is saved.
The flat data are taken when the telescope is in fast and random movement. The field of view (FOV) is kept in a quiet region near the disk center.
The average result of more than five hundred frames, $\emph{F}_{\rm mean}$, is used for the follow-up purpose of flat-fielding and calibration.
The science data was a set of sit-and-stare data acquired with the AO system, focused on the umbra of a sunspot from 10:27:38 UT to 10:33:14 UT.
In this paper, this set of data is used as an example to demonstrate the complete reduction process, which can be used in both the sit-and-stare and raster-scan spectrum modes.

\section{Data Processing} 
      \label{S-4}

In this section, we analyse the causes of non-uniformities, and describe the detailed steps of data processing, including distortion correction, flat-field correction, spectral line position alignment, intensity normalization, and wavelength calibration.
Before processing, all the science data and the $\emph{F}_{\rm mean}$ are subtracted with the $\emph{D}_{\rm mean}$.

\subsection{Distortion Correction} 
  \label{S-4.1}

The spectral distortion is mainly caused by the slight inaccuracy of the spectrometer system. For example, the CCD edges and the dispersion direction or spatial direction are not strictly parallel or vertical, which leads to inclination of the spectral lines. The deviation between the incident grating angle and the center of the grating causes curvature of the spectral lines. In order to acquire an accurate flat-field, these distortions have to be corrected. During the distortion correction process, the inclination is corrected by the horizontal black stripe that is caused by defects of the slit edge or dust attached to the slit. The curvature is corrected by the terrestrial lines. The horizontal stripe and the terrestrial lines are all from the $\emph{F}_{\rm mean}$.

      \begin{figure}    
   \centerline{\includegraphics[width=1.0\textwidth,clip=]{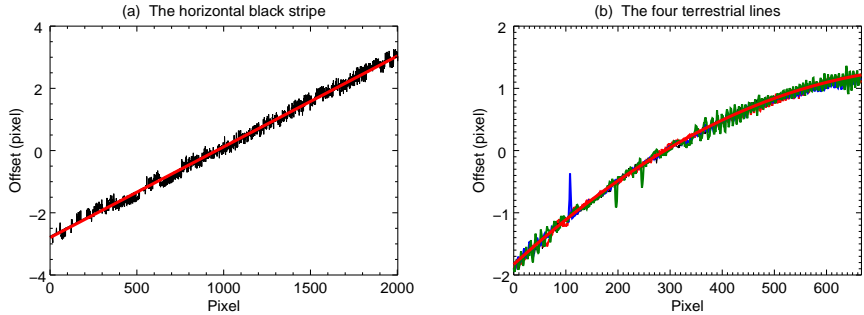}
              }
              \caption{(a) The relative positions (black) and the fitting curve (red) of the horizontal black line (wavelength direction). (b) The relative positions (with different color) and the fitting curve (red) of the four terrestrial lines (slit direction). The \emph{Y} axes gives the values of distortion (pixel units).
                      }
   \label{F-1}
   \end{figure}

\subsubsection{Correction of the X Inclination} 
  \label{S-4.1.1}
The deepest horizontal black stripe is chosen as the baseline for the horizontal inclination correction. The positions of the stripe in each column are calculated with a sub-pixel centroid algorithm. The median of these positions is used as the reference value, and the relative positions are obtained as shown in Figure~\ref{F-1}a. The figure shows that the curve of relative positions exhibits a linear trend. Then a linear fitting (red line in Figure~\ref{F-1}a) is used to get the inclination angle, which is 0.167$^{\circ}$. The corrected matrix $\emph{F}_{{\rm mean}X}$ is obtained by rotating $\emph{F}_{\rm mean}$ with the inclination angle. The incomplete rows at the top or the bottom of the image are discarded.

\subsubsection{De-stretching of the Y Curvature} 
  \label{S-4.1.2}
After the \emph{X}-direction inclination corrected, the curvature of spectral lines along the \emph{Y}-direction still exist. In order to accurately correct the curvature, the four terrestrial lines (6547.7\,{\AA}, 6548.6\,{\AA}, 6552.6\,{\AA}, 6572.1\,{\AA}) located on both sides of the H$\alpha$ line are selected to calculate the curvature offsets. By using the sub-pixel centroid algorithm, the positions of the four terrestrial lines in each row are determined.
The relative position curves in Figure~\ref{F-1}b show that the curvature of the four terrestrial lines are very similar and are  nonlinear. So a second-order polynomial approximation is used to fit the median of the four relative positions (red line in Figure~\ref{F-1}b), and then the curvature displacement of each row is obtained.
Each row of $\emph{F}_{{\rm mean}X}$ is shifted along the \emph{X}-direction with the opposite direction of its displacement.
Similarly, the incomplete columns are also cut away. The $\emph{F}_{{\rm mean}X}$ after curvature correcting is named as $\emph{F}_{{\rm mean}XY}$.

\subsection{Flat-fielding} 
  \label{S-4.2}
\subsubsection{Problems with the Flat-fielding} 
  \label{S-4.2.1}

The purpose of flat-fielding is to correct the non-uniformities of the system response, including the non-uniform sensitivity of the pixels across the chip of the CCD, and the non-uniform illumination. Comparing with the flat data of the imaging observation, the spectral flat data always contains not only non-uniformities, but also spectral lines. In order to obtain the correct flat-field, the spectral lines have to be removed from the $\emph{F}_{{\rm mean}XY}$. The traditional methods divide each row of $\emph{F}_{{\rm mean}XY}$ by the mean profile which is the average of all rows. In this paper, this matrix is defined as the primary flat-field. However, when the primary flat-field is used to process our spectrum, the following problems still exist that will reduce the precision of the physical parameters.

\begin{figure}    
   \centerline{\includegraphics[width=1.0\textwidth,clip=]{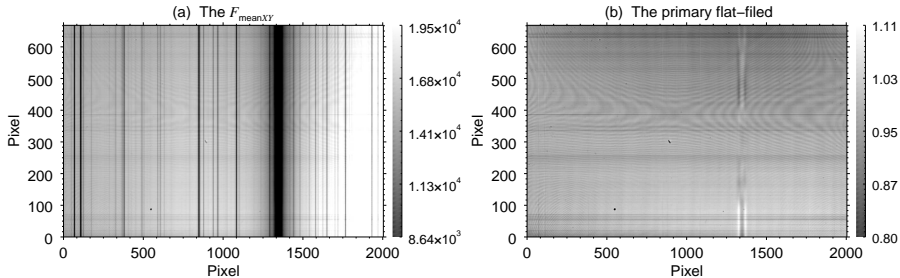}
              }
              \caption{(a) The $\emph{F}_{{\rm mean}XY}$, which is the average flat-field frame $\emph{F}_{\rm mean}$ after the distortion correction. (b) The primary flat-field. The grey-scale bar gives relative intensities expressed in ADU (Analog-Digital Units).
                      }
   \label{F-2}
   \end{figure}

The first problem is that the primary flat-field cannot be used to correct the horizontal continuum gradient of science data,
because the gradient in the primary flat-field is eliminated when each row of $\emph{F}_{{\rm mean}XY}$ is divided by the mean profile.
However, this gradient still exists in the science data.
The $\emph{F}_{{\rm mean}XY}$ in Figure~\ref{F-2}a shows the obvious intensity gradient of data. Figure~\ref{F-2}b shows the primary flat-field, in which the obvious intensity gradient is deducted.

The second problem is that remnants of spectral lines exist in the primary flat-field, which caused by the following reasons.
First, the solar structure in spectral lines is not completely smoothed, especially in the broad and deep H$\alpha$ spectral line, which can be clearly seen in Figure~\ref{F-2}b. Second, the solar differential rotation within the FOV results in the inclination of the solar spectral lines. As mentioned above, the four terrestrial lines are used to obtain the curvature displacements of the spectral line. But we find that even though all the modified terrestrial lines are almost parallel to the \emph{Y} axes, the modified photospheric lines still have a slight inclination. This phenomenon may be caused by solar differential rotation within the FOV. The magnitude of this inclination is related to the position of the slit on the solar surface during the observation.
The third reason is that although the curvature displacements of spectral lines are accurately calculated, there still exist a small amount of unavoidable deviation, and partial spectral lines are not deducted entirely.
These remnants in the obtained flat-field are not considered in traditional approaches, but they will introduce new errors when the flat-field is used for processing the science data.

The last problem is that there are two temporal changes in the primary flat-field, horizontal stripes and interference fringes.
The drift of horizontal stripes is caused by the spectrograph drift during the observations. This phenomenon is similar with the spectrograph of VTT (\citealp{VTT2002}).
Besides the drift of horizontal stripes, the interference fringes are the other time-varying factor in the NVST spectral data, especially in the infrared wavelengths. Moreover, the shapes of interference fringes are irregular, and the intensity and phase change with the incidence angle.
If the flat data are taken frequently, or the acquisition time of flat data is very close to that of science data, this phenomenon can be neglected.
In many cases, a long time and continuous observation is required to study the evolution of solar structure information, especially for raster-scan observations. The flat data could be taken only in the beginning and the end of the observations, in which case the flat-field conditions are changed.
In order to avoid the influences of the time-varying factors, they should be extracted from the full flat-field and further corrected before being used to process the science data.

\subsubsection{Extraction and Correction of the Non-uniform Matrices} 
  \label{S-4.2.2}
\begin{figure}    
   \centerline{\includegraphics[width=1.0\textwidth,clip=]{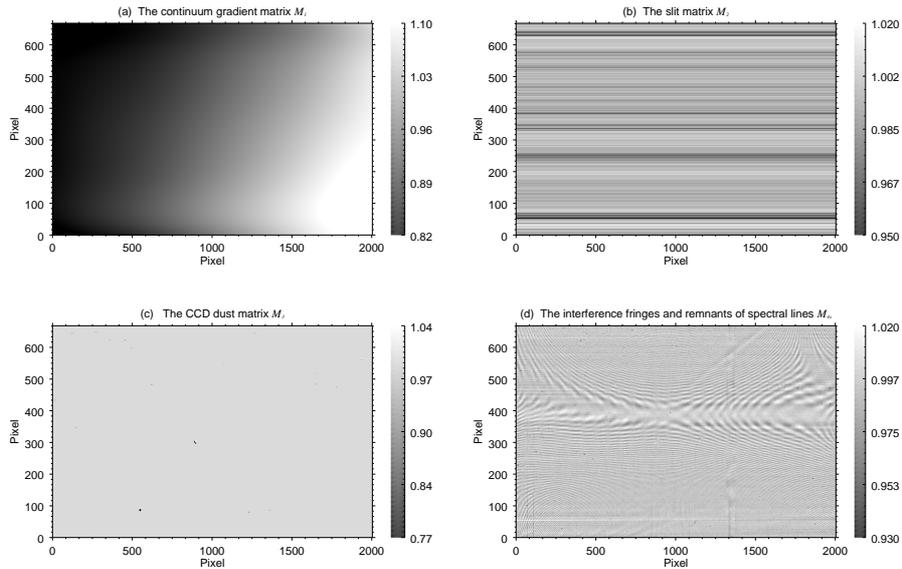}
              }
              \caption{(a) The continuum gradient matrix $\emph{M}_{1}$. (b) The matrix of the slit non-uniformity matrix $\emph{M}_{2}$. (c) The CCD dust matrix $\emph{M}_{3}$. (d) The interference fringes and remnants of spectral lines $\emph{M}_{4s}$. The grey-scale bars give relative intensities normalized to 1.0.}
   \label{F-3}
   \end{figure}

Based on the above three problems, if the primary flat-field is used to process the science data, not only some non-uniformities still remain in the result, but also a lot of extra false signals are introduced. So we need to make some improvements to the traditional approaches. The best method to solve those problems is to separate the full flat pattern into several independent matrices and then deduct them from the data one by one. According to the characteristic of non-uniformities in the NVST spectra, the ideal flat-field is defined with the following equation,
 \begin{equation}  \label{Eq-Flat}
    \emph{F} =  \emph{M}_{1} \cdot \emph{M}_{2} \cdot \emph{M}_{3} \cdot \emph{M}_{4} \,.
  \end{equation}
Where \emph{F} denotes the full flat-field matrix, $\emph{M}_{1}$ is the continuum gradient matrix, $\emph{M}_{2}$ is the slit non-uniformity matrix, $\emph{M}_{3}$ is the CCD dust matrix, and $\emph{M}_{4}$ is the interference fringe matrix.
In order to keep the continuum gradient when removing spectral lines from the flat-field, $\emph{F}_{{\rm mean}XY}$ is divided by a new profile, which is normalized by fitting a curve to the continuum of the mean profile. The result, defined as $\emph{F}_{\rm extend}$, contains $\emph{M}_{1}$, $\emph{M}_{2}$, $\emph{M}_{3}$, $\emph{M}_{4}$ and spectral lines remnants. To removed the time-varying factors from the flat-field, $\emph{M}_{2}$ and $\emph{M}_{4}$ should be extracted from $\emph{F}_{\rm extend}$. Since $\emph{M}_{2}$ has a relatively stable shape and just drifts along the \emph{Y} axes (horizontal drift can be ignored), it can be directly extracted from $\emph{F}_{\rm extend}$. $\emph{M}_{4}$ is very difficult to extract directly from $\emph{F}_{\rm extend}$, because its shape is irregular, and its intensity and phase change with the incidence angle. To solve this problem, the method used in this paper is to extract $\emph{M}_{1}$, $\emph{M}_{2}$ and $\emph{M}_{3}$ respectively, and leave the remnants of spectral lines and the $\emph{M}_{4}$ behind.
To avoid the errors caused by the curve fitting with the continuum in $\emph{F}_{\rm extend}$, $\emph{F}_{{\rm mean}XY}$ is used as the extracted matrix instead of $\emph{F}_{\rm extend}$. The detailed steps are the following:

The continuum gradient is mainly caused by the non-uniform illumination and stray light of optical path.
considering the physical parameters, like spectral line symmetry, are affected by the continuum gradient, the following procedure is used to correct $\emph{M}_{1}$.
At first, the $\emph{F}_{{\rm mean}XY}$ is normalized.
Then, the regions of normalized $\emph{F}_{{\rm mean}XY}$, where only contain continuum, are used to fit the continuum gradient matrix $\emph{M}_{1}$ with a fifth-order surface polynomial.
The fitting result is shown in Figure~\ref{F-3}a.
A new matrix is named as $\emph{F}_{1}$, which is the result of dividing $\emph{F}_{{\rm mean}XY}$ by $\emph{M}_{1}$.

The slit non-uniformities appear as horizontal black lines and bright lines in data, which are caused by imperfection of the slit as mentioned above. The average of $\emph{F}_{1}$ along the \emph{X}-direction can be used as the slit pattern. However, considering that CCD dust and interference fringes may bring some systematic errors, the median of all the columns of $\emph{F}_{1}$ is used as the slit profile. Then, this profile is replicated along the \emph{X}-direction to get the slit non-uniformity matrix $\emph{M}_{2}$ which is shown in Figure~\ref{F-3}b.
The matrix $\emph{F}_{2}$ is the result of $\emph{F}_{1}$ divided by $\emph{M}_{2}$.

The CCD dust matrix is caused by the dust particles lying on the CCD surface.
A morphological method is used in extracting $\emph{M}_{3}$. For a better extraction result, the intensity distribution of the extracted matrix needs to be relatively uniform except $\emph{M}_{3}$, while the matrix $\emph{F}_{2}$ still contains spectral lines, CCD dust and interference fringes.
To remove the spectral lines, each row of $\emph{F}_{2}$ is divided by the mean profile, and the result is defined as $\emph{F}_{3}$. After this process, some remnants of spectral lines, as well as the interference fringes and CCD dust, remain in $\emph{F}_{3}$. So we have to smooth these remnants and fringes by following two steps: The first step is that a large scale (40 pixel) median-filter processing is adopted for each column of $\emph{F}_{3}$ to obtain a new matrix, and $\emph{F}_{3}$ is divided by this matrix to give the matrix \emph{S}. The second step is to carry out the same processing for each row of \emph{S}, and the smoothed matrix $\emph{F}_{3s}$ is obtained, which just contains the CCD dust and a small amount of interference fringes.
To extract the CCD dust matrix from $\emph{F}_{3s}$, a proper threshold (the mean minus 5 times the standard deviation) is taken for $\emph{F}_{3s}$ to get a binary matrix, in which the values less than the threshold are set to 1 and the others are 0. Then, the binary matrix is subject to morphological dilation with a 5 $\times$ 5 pixels$^{2}$ structure to restore the shape of CCD dust. The CCD dust gray matrix is the restored binary matrix multiplied by $\emph{F}_{2}$. In this gray matrix, the values of CCD dust intensities are restored. The CCD dust matrix $\emph{M}_{3}$ is obtained by keeping the values of CCD dust in the gray matrix and setting others values to 1.0, and the result is shown in Figure~\ref{F-3}c.
After $\emph{F}_{3}$ is divided by $\emph{M}_{3}$, the result is $\emph{M}_{4s}$, in which the interference fringes and remnants of spectral lines are left. This is displayed in Figure~\ref{F-3}d.

Based on the methods explained above, we get four independent flat matrices ($\emph{M}_{1}$, $\emph{M}_{2}$, $\emph{M}_{3}$ and $\emph{M}_{4s}$).
These matrices are extracted from $\emph{F}_{{\rm mean}XY}$, with the geometric distortion in both horizontal and vertical directions corrected.
Thus, before flat-fielding, the science data should be processed the same geometric distortion as $\emph{F}_{\rm mean}$ to align with the four flat-field matrixes.
After the geometric correction, $\emph{M}_{1}$ and $\emph{M}_{3}$, which are relatively stable, can be directly divided from the science data.
However, $\emph{M}_{2}$ is changed in the science data as described in Section~\ref{S-4.2.1}.
In order to correct the slit non-uniformity completely, the deepest horizontal strips which exist in both $\emph{F}_{\rm mean}$ and the science data are chosen to calculate their positions by the sub-pixel centroid algorithm.
The relative displacements of $\emph{M}_{2}$ are obtained by using the position of $\emph{F}_{\rm mean}$ as a reference value.
Then, $\emph{M}_{2}$ is modified by vertical shifting with the opposite relative displacements, respectively.
After that, the slit non-uniformity in the science data can be eliminated accurately by using the shifted $\emph{M}_{2}$.
In principle, further steps should be made to process the interference fringes. However, because of the particularity of interference fringes as described in Section~\ref{S-4.2.1}, it is very difficult to extract the pure $\emph{M}_{4}$ from the flat-field, even from the science data.
So we temporarily do not deal with the interference fringes. Their impacts will be analyzed in Section~\ref{S-5}.

\subsection{Additional Correction to the Science Data} 
  \label{S-4.3}

After flat-fielding, there still exist two additional temporal changes in the science data: the position drift of spectral lines and relative intensity fluctuation.
The position drift of spectral lines is caused by the spectrometer drift, which also brings the above slit non-uniformity matrix drift.
The correction of this drift is to align the wavelength of each data frame, which helps to study the evolution of physical parameters at a fixed spatial position (like solar oscillation, \emph{etc}.), and prepares for the future processing of raster-scan spectra (for example, ensuring the monochromaticity of the raster images composed by the science data).
The correction of intensity fluctuation is important for long-period observations. It is useful for comparing the physical parameters between frames and for increasing the spatial resolution of the raster images.

For most sit-and-stare spectral data, the correction of spectral line drift and relative intensity fluctuation could be ignored, and the drift correction could be applied in deriving the spectral parameters associated with the positions of spectral lines. But to build the universal procedure for sit-and-stare and raster-scan spectrum and to obtain more standard three-dimensional spectral arrays ($\emph{X}$, $\emph{Y}$, wavelength), both corrections are applied to science data.

\subsubsection{Spectral Line Position Aligning} 
  \label{S-4.3.1}

\begin{figure}    
   \centerline{\includegraphics[width=1.0\textwidth,clip=]{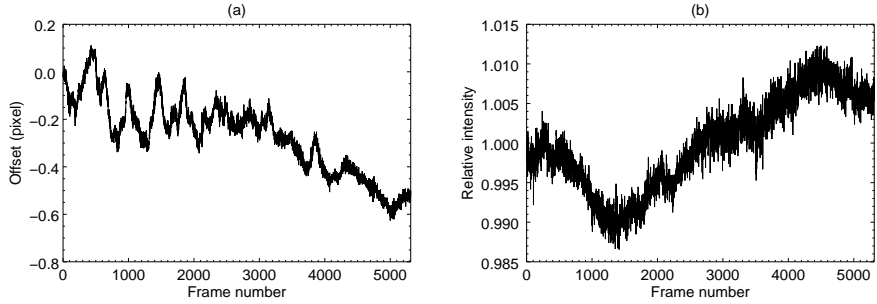}
              }
              \caption{(a) The drift offsets of the deepest terrestrial line over the observation time. The \emph{Y} axes gives the value of the relative position. (b) The relative intensity distribution of the limited regions over the observation time. The \emph{Y} axes gives the relative intensity.
                      }
   \label{F-4}
   \end{figure}

To correct the drift of spectral lines, the positions of the deepest terrestrial lines in $\emph{F}_{\rm mean}$ and the science data are calculated by the sub-pixel centroid algorithm.  The position of $\emph{F}_{\rm mean}$ is used as the reference value to obtain the drift offsets, which are shown in Figure~\ref{F-4}a.
This figure shows obvious fluctuation of the position, and the ranges are about -0.2\,--\,0.5 pixels corresponding to the Doppler velocities of -220\,--\,550\,m\,s\textsuperscript{-1}.
The science data are shifted horizontally with the inverse drift offsets, and this ensures consistency of spectral lines.

\subsubsection{Intensity Normalization} 
 \label{S-4.3.2}

In science data, the average intensities in the same limited region (the rectangular box in Figure~\ref{F-6}a) in each frame are calculated.
Then, the median of all the average intensities is used as the reference value, and the normalization factors are obtained by the average intensity of each limited region divided by the reference value.
The relative intensity distribution of the limited region is shown in Figure~\ref{F-4}b.
Because the data used only last about five minutes, the fluctuations of the intensity distribution in Figure~\ref{F-4}b are not obvious.
After the correction of intensity fluctuations, the science data all are normalized by the normalization factors.

\subsection{Wavelength Calibration} 
  \label{S-4.4}

\begin{figure}    
   \centerline{\includegraphics[width=1.0\textwidth,clip=]{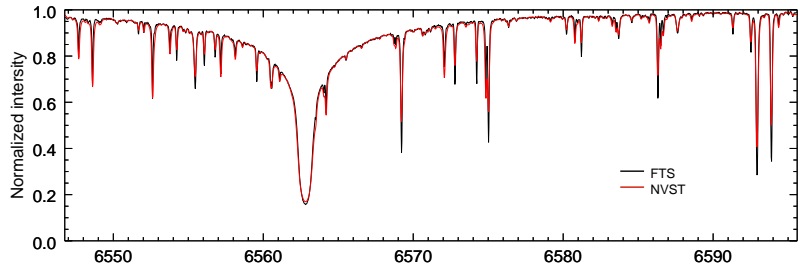}
              }
              \caption{The calibration results between the NVST spectral profile (red line) and FTS spectrum (black line). The NVST spectral profile is the average of the reduced $\emph{F}_{\rm mean}$ along the \emph{Y}-direction. The \emph{Y} axes gives relative intensities normalized to 1.0.
                      }
   \label{F-5}
   \end{figure}

To acquire precise wavelengths of the spectral lines and the spectral sampling rate of the spectrum, wavelength calibration is applied after the above processing.
Because the position of spectral lines in flat data and science data are aligned, wavelength calibration can be obtained by the reduced $\emph{F}_{\rm mean}$.
The reference spectral lines for wavelength calibration need to be as stable as possible, and the terrestrial lines are usually a good choice.
The positions of four terrestrial lines (the same lines used for geometric correction)  in the NVST spectrum are firstly calculated.
Meanwhile, the wavelengths of the four terrestrial lines are obtained from the standard solar spectrum atlas which is observed by the \emph{Fourier Transform Spectrometer} (FTS) at the \emph{McMath-Pierce Solar Telescope}.
The wavelength calibration is applied by fitting the positions found in NVST spectrum and wavelengths obtained from FTS with the least square method.
The spectral sampling rate of the NVST spectrum is also acquired, which is 0.0244 $\pm$ 0.0001\,{\AA} pixel\textsuperscript{-1}. Figure~\ref{F-5} shows a comparison result between the NVST reduced $\emph{F}_{\rm mean}$ profile (red line) and the FTS spectrum (black line).

\section{Results and Discussion} 
      \label{S-5}
\begin{figure}    
   \centerline{\includegraphics[width=1.0\textwidth,clip=]{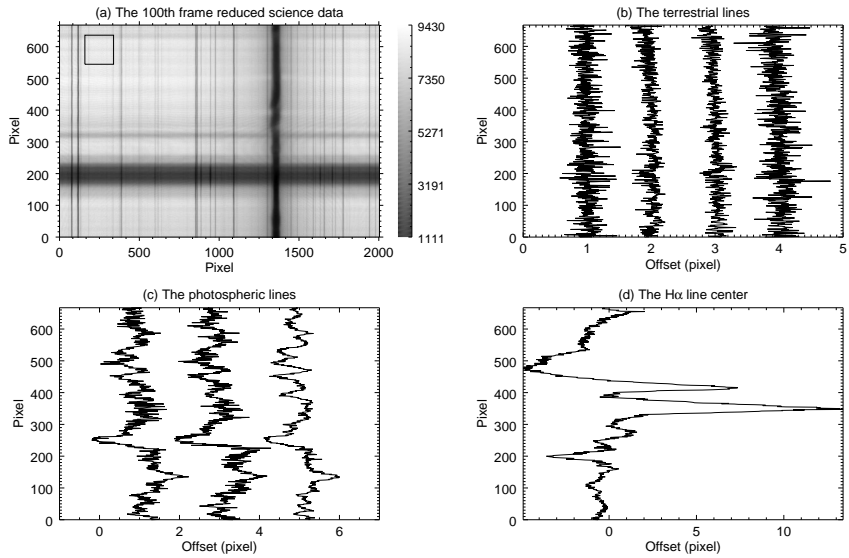}
              }
              \caption{ The results of the reduced science data. (a) The 100th frame science data after the precise processing. (b) The Doppler displacement curves of the four terrestrial lines. (c) The Doppler displacement curves of three photospheric lines. (d) The Doppler displacement curve of the H$\alpha$ line center. In order to make the results clearer, we set the interval between lines in the panel (b) as 1.0 pixel and in panel (c) as 2.0 pixels.
                      }
   \label{F-6}
   \end{figure}

This section shows the result of fine processed science data. The measuring accuracies of the Doppler replacements are use to evaluate whether the reduced NVST spectrum can reach the scientific requirement. Here the Doppler replacements are the same as the above relative positions of spectral lines, calculated by the sub-pixel centroid algorithm.
Figure~\ref{F-6}a shows the reduced 100th frame science data. It shows that the continuum gradient, the slit non-uniformity and the CCD dust are well deducted.
The relative standard deviations of the rectangular regions, mentioned in Section~\ref{S-4.3.2}, in continuum area of raw data and reduced data are 2.1\% and 1.4\%, respectively. This means the noise is well suppressed after fine processing. Figure~\ref{F-6}b shows the Doppler displacement curves of four terrestrial lines. As can be seen from the figure the geometric curvature of spectral lines is well corrected.
The Doppler displacement curves of three other photospheric lines, shown in Figure~\ref{F-6}c, shows a strong correlation.
The average shift range is about -1.24\,--\,1.21 pixels corresponding to -1.36\,--\,1.33\,km\,s\textsuperscript{-1} Doppler velocities. That reflects the motion characteristics of the solar photosphere at this moment.
Figure~\ref{F-6}d shows the Doppler displacement of the H$\alpha$ line center, and the range is -5.41\,--\,14.73\,km\,s\textsuperscript{-1}.

In the reduced data, the non-uniformities of continuum gradient, slit and CCD dust are all corrected except the interference fringes. Here, we will analyze the influence of the interference fringes on the accuracy of our measurements.
The $\emph{F}_{\rm mean}$ is the averaged result of the multi-frame flat data, and each of the flat data was acquired during the random movement of the telescope, so the fine solar structures in the FOV of the slit are smoothed. The root mean square (RMS) of the Doppler displacements in the reduced $\emph{F}_{\rm mean}$ mainly reflects the impact of the interference fringes.
The RMS of the seven spectral lines are listed in Table~\ref{T-3}, and the range is 0.021\,--\,0.038 pixels, corresponding to Doppler velocities of 23.1\,--\,41.8\,m\,s\textsuperscript{-1}.
Of course, due to the time-varying property of the interference fringes, it is hard to calculate the interference fringe influence exactly at each position of data. In some positions, the impact may reach up to 50\,m\,s\textsuperscript{-1}, or even greater.

\begin{table}
\caption{The RMS of four terrestrial lines and three photospheric lines in the reduced $\emph{F}_{\rm mean}$.}.
\label{T-3}
\begin{tabular}{lllll}     
   \hline              
  Terrestrial lines ({\AA}) & 6572.1   & 6552.6 & 6548.6  & 6547.7 \\
   RMS & 0.032 & 0.038  &  0.035  &  0.023 \\
  \hline            
  Photospheric lines ({\AA}) &6554.2  & 6556.0  & 6569.2  \\
   RMS               & 0.036    &  0.021   &   0.033 \\
   \hline
\end{tabular}
\end{table}

The final measuring accuracy of reduced science data is our greater concern.
In order to avoid the influence of the solar structures on the measuring accuracy calculation, the measuring accuracy is inferred by a set of second-order finite differences, which are calculated from three neighboring pixels.
The depth of the spectral line has a great influence on the measuring accuracy of the Doppler velocity.
Therefore, the measuring accuracy and the depth of the terrestrial lines and the photospheric lines are calculated respectively, and the result are listed in Table~\ref{T-4}.
It shows that there is a strong inverse correlation between the depth and the measuring accuracy in both terrestrial lines and photospheric lines. The correlation coefficients are -0.98 and -0.97, respectively.
Namely the deeper the spectral lines, the higher the accuracy of Doppler velocity measurement.
In Table~\ref{T-4}, it is also shown that when the depth of photospheric lines is within 0.314\,--\,0.615, the measuring accuracy can reach up to 71.5\,--\,157.3 m\,s\textsuperscript{-1}.
In the H$\alpha$ spectral line center, the measuring accuracy is 0.121 pixels, corresponding to 133 m\,s\textsuperscript{-1}.

\begin{table}
\caption{The depth and measuring accuracy of four terrestrial lines and three photospheric lines in the reduced science data. }.
\label{T-4}
\begin{tabular}{lllll}     
  \hline              
  Terrestrial lines ({\AA}) & 6572.1   & 6552.6   & 6548.6   & 6547.7  \\
   Depth                &  0.241  &  0.381  &  0.323  &  0.200  \\
   Accuracy (m\,s\textsuperscript{-1}) &  156.2   & 75.9    &  91.3  &  172.7 \\
  \hline            
  Photospheric lines ({\AA})  &6554.2  & 6556.0  & 6569.2  \\
   Depth               & 0.314    &  0.317   &   0.615 \\
   Accuracy (m\,s\textsuperscript{-1})  &  136.4   & 157.3   &  71.5 \\
   \hline
\end{tabular}
\end{table}

\begin{figure}    
   \centerline{\includegraphics[width=1.0\textwidth,clip=]{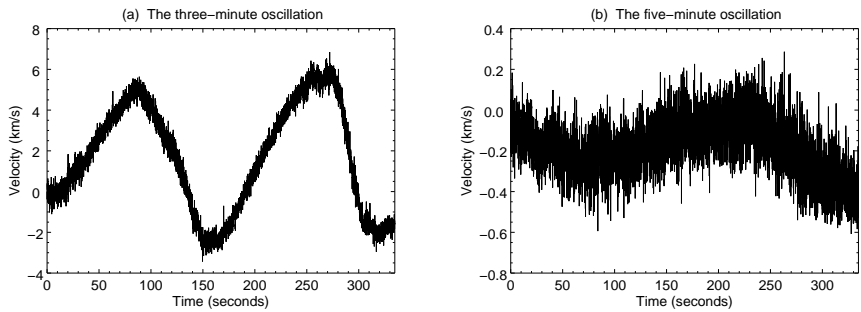}
              }
              \caption{(a) The Doppler velocity distribution of the H$\alpha$ line center in the sunspot umbra region, which present the three-minute oscillation of the solar chromosphere. (b) The Doppler velocity distribution of the photospheric quiet region, which is the five-minute oscillation of the solar photosphere. The \emph{X} axes gives the acquisition time (in seconds), and the \emph{Y} axes gives the Doppler relative velocity (in m\,s\textsuperscript{-1}).
                      }
   \label{F-7}
   \end{figure}

In addition, the Doppler velocity of the H$\alpha$ line center in the sunspot umbra region of all the reduced science data is also obtained, and the result is shown in Figure~\ref{F-7}a.
This figure shows the three-minute oscillation of the solar chromosphere, and the range of the fluctuation is -5.5\,--\,3.0 km\,s\textsuperscript{-1}.
Figure~\ref{F-7}b shows the Doppler velocity evolution curve of a quiet region in the photosphere. It exhibits an oscillation with a cycle of five minutes, and the range of this five-minute oscillation is -0.5\,--\,0.1 km\,s\textsuperscript{-1}.
The above results are consistent with the basic facts of solar physics. The measuring accuracies of the two oscillation curves are 0.186 pixels and 0.069 pixels, corresponding to 204.6 and 75.9 m\,s\textsuperscript{-1}, respectively.

In conclusion, the results shows by this precise reduction that the highest Doppler measuring accuracy of photospheric lines that can be achieved is within 100 m\,s\textsuperscript{-1}, and the Doppler measuring accuracy of the H$\alpha$ spectral line is about 100\,--\,200 m\,s\textsuperscript{-1}.
The actual measuring accuracies are always affected by systematic errors and random noise. In this paper, the corrected non-uniformities are systematic errors. The interference fringes and random noise needs to be further suppressed.

\section{Conclusion} 
      \label{S-6}

In this paper, the precise reduction methods and procedure presented can effectively correct particular non-uniformities of the sit-and-stare and raster-scan spectra of NVST.
The separation method is used to do an accurate flat-fielding, which can not only reduce most of the influences of the non-uniformities, but also avoid introducing additional errors.
The additional corrections are applied to the science data to obtain a standard three-dimensional reduced spectral arrays.
After the fine processing, the measuring accuracies of Doppler velocities with different spectral lines, and of the curves of the chromospheric three-minute oscillation and photospheric five-minute oscillation, demonstrate the validity of this procedure.
The Doppler velocity measuring accuracy of the H$\alpha$ line center can reach up to 100\,--\,200\,m\,s\textsuperscript{-1}.
The accuracies of photospheric lines depend on their line depths, and the highest Doppler measuring accuracy is within 100\,m\,s\textsuperscript{-1}.
This shows that by this fine processed spectrum, the characteristics of the chromosphere and the photosphere can be studied with reduced NVST data co-spatially and co-temporally.
Some work remains to do, like removing interference fringes and image de-noising, and this will be studied in future work.

\begin{acks}
We are appreciated for all the help from the colleagues in the NVST team. We are also thankful to the unknown referee for their useful comments.
A lot of thanks to Mr. Song Feng, Ms. Yanxiao Liu and Dr. Huanwen Peng for their kind assistance and helpful comments on this manuscript.
The FTS atlas used in this paper were produced by the NSO/NOAO.
This work is supported by the National Natural Science Foundation of China (NSFC) under grant numbers 11773072, 11573012 and 11473064.
\end{acks}

\medskip

\noindent {\footnotesize \textbf{Disclosure of Potential Conflicts of Interest} \quad The authors declare that they have no conflicts of interest.}


\end{article}
\end{document}